\documentclass[aps,pra,amsmath,amssymb,showpacs,twocolumn]{revtex4}

\usepackage{amssymb}
\usepackage{amsmath}
\usepackage{amscd}
\usepackage{natbib}
\usepackage{graphicx}            

\newcommand{\be}{\begin{equation}}
\newcommand{\ee}{\end{equation}}

\renewcommand{\vec}[1]{{\mathbf #1}}

\begin{document}


\title{Equation of state and Kosterlitz-Thouless transition temperature in two-dimensional Fermi gases: An analytical approach}

 \author{Michael Klawunn}
 \affiliation{Institut f\"ur Theoretische Physik, Leibniz Universit\"at Hannover, Appelstrasse 2, 30169, Hannover, Germany}

\date{\today}

\begin{abstract}
We study Fermi gases in two dimensions at low temperatures with attractive interactions.
Analytic results are derived for the equation of state and the Kosterlitz-Thouless transition temperature as functions of the two-body binding energy and the density of the gas. Our results for the equation of state and the pressure of the gas strongly deviate from the mean field predictions. However, they are in reasonable agreement with Monte-Carlo calculations and recent experiments with cold atomic gases.
\end{abstract}


\maketitle
\section{Introduction}

In the last years ultracold two dimensional Fermi gases have attracted a lot of interest. One reason is the assumption that pairing of fermions in two dimensions can help to explain unconventional superconductivity. 
Already in the 1980s, the two dimensional Fermi gas has been studied theoretically with a mean field approach based on the famous BCS theory and the BCS-BEC crossover in three dimensions \cite{Miyake1983,Randeria1990}.
In 3D, the BCS-state evolves to a superfluid BEC of tightly bound molecules by tuning the scattering length from negative to positive values.

The BCS-BEC crossover in two dimensions differs fundamentally from the 3D case. 
In 3D, in the BCS-regime, fermionic pairing is governed by a many-body order parameter of Cooper pairs, while a two-body bound state is lacking. Contrary, in 2D even in the BCS-regime, a two body bound state exists, which significantly influences the physics of the system.  
In the BEC-regime in 3D, the tightly bound molecular pairs form the order parameter of a superfluid BEC of molecules. In 2D, a molecular BEC can strictly form only at T=0.
However, a superfluid with an algebraically decaying correlation function can form at finite temperatures below the Kosterlitz-Thouless (KT) transition temperature \cite{KT}. The superfluid order parameter follows from the phase of the wave-function of the molecular gas and the appearance of vortex-antivortex pairs.
Although in 2D fermionic pairing and the BCS-BEC crossover are fundamentally different, we keep using the expressions BCS and BEC for comparisons and discussions throughout the paper.  
The evolution of fermion pairing from three to two dimensions has been investigated experimentally in \cite{Sommer2012}.

The equation of state at zero temperature for the BCS-BEC crossover in two dimensions has been determined with a fixed-node variational Monte-Carlo calculation \cite{Bertaina2011}. The results deviate even qualitatively from the mean field approximations. 
Two recent experiments with trapped cold atoms are in good agreement with the Monte-Carlo calculations \cite{Makhalov2014, Ong2015}. The BCS-BEC crossover in 2D has been also investigated recently in \cite{Boettcher2015} experimentally and with a Luttinger Ward theory. Further recent theoretical studies on the equation of state in the 2D BCS-BEC crossover are \cite{Salasnich2015,He2015,Shi2015}. A Monte-Carlo calculation for finite temperature is given in \cite{Drut2015}.
Additionally, spin imbalanced Fermi gases in two dimensions has been investigated in \cite{Ong2015}. 

The KT-transition temperature for a Fermi gas has been calculated in \cite{Sademelo2006}. In the molecular limit the KT-temperature should converge to the KT-temperature of the weakly interacting Bose gas \cite{Petrov2003, Prokofiev2001}.
Recently, the equation of state at finite temperatures and the KT-transition temperature have been obtained numerically with a Luttinger-Ward model \cite{Bauer2014} and with Gaussian fluctuations beyond mean field \cite{Bighin2015}. Experimentally, the KT condensate has been investigated in \cite{Ries2015}.     

However, the derivation of an analytical expression for the equation of state, which is qualitatively valid through the whole crossover, is still lacking. From such an expression one can simply derive thermodynamic quantities like the chemical potential or the pressure of the gas.
Therefore, the goal of this paper is to provide an analytical model for the two dimensional Fermi gas with attractive interactions. We derive simple expressions for the equation of state at zero temperature and the KT transition temperature as functions of the two-body binding energy $\epsilon_B$ and the density of the gas $n$. We validate the model by comparison with Monte-Carlo calculations \cite{Bertaina2011} and cold atom experiments \cite{Makhalov2014, Ong2015}.

\section{Equation of state}

As stated in the introduction, the mechanism of fermionic pairing is fundamentally different in a two dimensional system. In 2D, a many body system is significantly influenced by the two-body physics, and in particular the two-body bound state. Therefore, we choose an approach based on two-body scattering, where many body effects only enter approximately by excluding all intermediate states below the Fermi surface. This approximation is implemented in the Bethe-Goldstone integral equation \cite{BG1957}. It is known from nuclear physics \cite{Bethe1956, Brueckner1956, Bethe1962} and has been successfully used in the context of Brueckner-Hartree-Fock theory in condensed matter systems \cite{Lipparini}, e. g. for the Fermi-Polaron \cite{Alessio2007, Klawunn2011, Klawunn2013} and the 2D electron gas \cite{Nagano1984}. 

We consider a Fermi gas with an equal number $n/2$ of two sorts of particles, $|\uparrow >$ and
$|\downarrow >$, interacting attractively at zero temperature. 
The Bethe-Goldstone integral equation for the reaction matrix $G$ can be written as \cite{Bethe1962}
\begin{align}\label{BG}
\begin{split}
&G(\vec{k_i},\vec{P},\vec{k_f})=V(\vec{k_f}-\vec{k_i})+\int \frac{d\vec{k}}{(2\pi)^D}
V(|\vec{k_f}-\vec{k}|)\times\\
& \frac{Q(\vec{k})}
{E_i(\vec{P},\vec{k_i})
-E(\vec{P},\vec{k})}
G(\vec{k},\vec{k_i},\vec{P})
\end{split}
\end{align}
Here $\vec{k_i}$, $\vec{k}$ and $\vec{k_f}$ and are the initial, intermediate and final relative momentum of the two interacting particles. $E_i(\vec{P},\vec{k_i})$ and $E(\vec{P},\vec{k})$ are the initial and intermediate energies.
$\vec{P}$ is half of their center of mass momentum and $V(\vec{k})$ is the Fourier transform
of the two-particle interaction potential. $Q(\vec{k})$ is the Pauli operator, which guarantees that the momentum of intermediate states lies above the Fermi surface. If $|\vec{P}+\vec{k}|>k_F$ and $|\vec{P}-\vec{k}|>k_F$, then $Q=1$ otherwise $Q=0$. Here $k_F=\sqrt{2\pi n}$ is the Fermi momentum and $n$ is the total density of the gas.
In Eq. (\ref{BG}) only ladder diagrams are summed. 

The interaction energy follows from the mean value of the reaction matrix $\epsilon_{int}= \left\langle G(\vec{k_i},\vec{P},\vec{k_f})\right\rangle$. 
The energy per particle is then given by $E/N=E_{FG}/N+\epsilon_{int}/n$, where $E_{FG}/N=\epsilon_F/2$ is the kinetic energy of the ideal Fermi gas and $\epsilon_F=k_F^2/2m$ is the Fermi energy.  

The interaction is characterized by an attractive short-range potential of arbitrary shape.
We can express the potential in terms of the two-particle scattering amplitude $f$, which obeys an integral equation similar to Eq. (\ref{BG}), but with $Q=1$ (Lippmann-Schwinger equation).
Replacing the potential by $f$ renormalizes Eq. (\ref{BG}) with respect to ultraviolet divergences and we obtain \cite{Abrikosov-book}
\begin{align}
\begin{split}\label{BG-f1}
&G(\vec{k_i},\vec{k_f},\vec{P})=f(\vec{k_i},\vec{k_f})+\phantom{\int}\\
\int\!\! \frac{d\vec{k}}{(2\pi)^D}
& \frac{f(\vec{k_i}, \vec{k_f}-\vec{k} ) \left(Q(\vec{k})-1\right) }
{E_i(\vec{P},\vec{k_i})
-E(\vec{P},\vec{k})}
g(\vec{k_1},\vec{k_2},\vec{k}),
\end{split}
\end{align}
where $f(\vec{k_i},\vec{k_f})$ is the off-shell scattering amplitude.
Typically, the momentum transfer is small $\vec{k_i} \approx \vec{k_f}$ and the main contribution
from the integral comes from small values of $\vec{k}$. Thus, we can
approximate the off-shell scattering amplitude by the on-shell scattering amplitude $f(\vec{k_i})$. 

Then the reaction matrix does not depend on the final momentum $\vec{k_f}$ and Eq. (\ref{BG-f1}) can be written as
\begin{align}\label{G}
\frac{1}{G(\vec{k_i})}\approx \frac{1}{f(\vec{k_i})}
-\int\!\! \frac{d\vec{k}}{(2\pi)^D}
& \frac{ Q(\vec{k})-1 }
{E_i(\vec{P},\vec{k_i})
-E(\vec{P},\vec{k})}
\end{align}
We further assume that initially every two Fermions with opposite momenta are paired on the two-body level with binding energy $\epsilon_B<0$ \cite{eb} and $P=0$. 
Then the energy denominator simply reduces to $E_i(\vec{P},\vec{k_i})-E(\vec{P},\vec{k}) \approx \epsilon_B - k^2/m$.
For the case of a bound state the scattering amplitude has a pole, such that
$f\left( k_i=\sqrt{m\epsilon_b}\right)^{-1}=0$.
The Pauli operator $Q(\vec{k})-1$ limits the integral to $k<k_F$, which reflects the effect of the background Fermi gas on the two body scattering. The reaction matrix in 2D is then simply given by
\begin{align}\label{G-bound}
\frac{1}{G(\epsilon_b)} \approx - \int_{0}^{k_F} \frac{dk }{2\pi}
 \frac{k}{|\epsilon_b|+\frac{k^2}{m}}
 =-\frac{m}{4\pi} \ln\left[1+\frac{2\epsilon_F}{|\epsilon_b|} \right]. 
\end{align}
The interaction energy follows from the mean value of the reaction matrix which reduces at T=0 to
$\epsilon_{int} = \int_{k_i < k_F}\!\! \frac{d\vec{k_i}}{(2\pi)^2}\int_{k_f<k_F}\!\! \frac{d\vec{k_f}}{(2\pi)^2}\, G(\epsilon_b,\epsilon_F)$.
This leads with Eq. (\ref{G-bound}) to an analytical formula for the equation of state \cite{note_imbalance}
\begin{align}\label{E}
\frac{E}{N} \approx \frac{\epsilon_F}{2} - \frac{\epsilon_F}{\ln\left[1+\frac{2\epsilon_F}{|\epsilon_b|} \right]}.
\end{align}

In the limit of weak interactions, i.~e.
$|\epsilon_b| \to 0$, we have, as expected, $E \to E_{FG}$.
In the opposite limit, $|\epsilon_b| \gg \epsilon_F$, 
we obtain the interaction energy $\epsilon_{int} \approx \epsilon_B/2 -\epsilon_F/2$. 
Thus, the total energy per particle converges to half of the molecular binding energy $E/N \approx \epsilon_B/2$ \cite{noteD}.
This is the expected result, because in this regime the gas consists of strongly bound molecular bosonic pairs. The fermonic character and hence the density dependence vanish.
Since our approach starts from two-body scattering of the fermions, the interaction between bosonic pairs is lacking. However, our theory shows, that the systems behaves completely bosonic for $|\epsilon_b| \gg \epsilon_F$. Thus, in the molecular limit the theory of weakly interacting Bose gases should be used \cite{Bertaina2011}.

\vspace{0.5cm}
 
\begin{figure}[ht]
\begin{center}
\includegraphics[width=0.44\textwidth,angle=0]{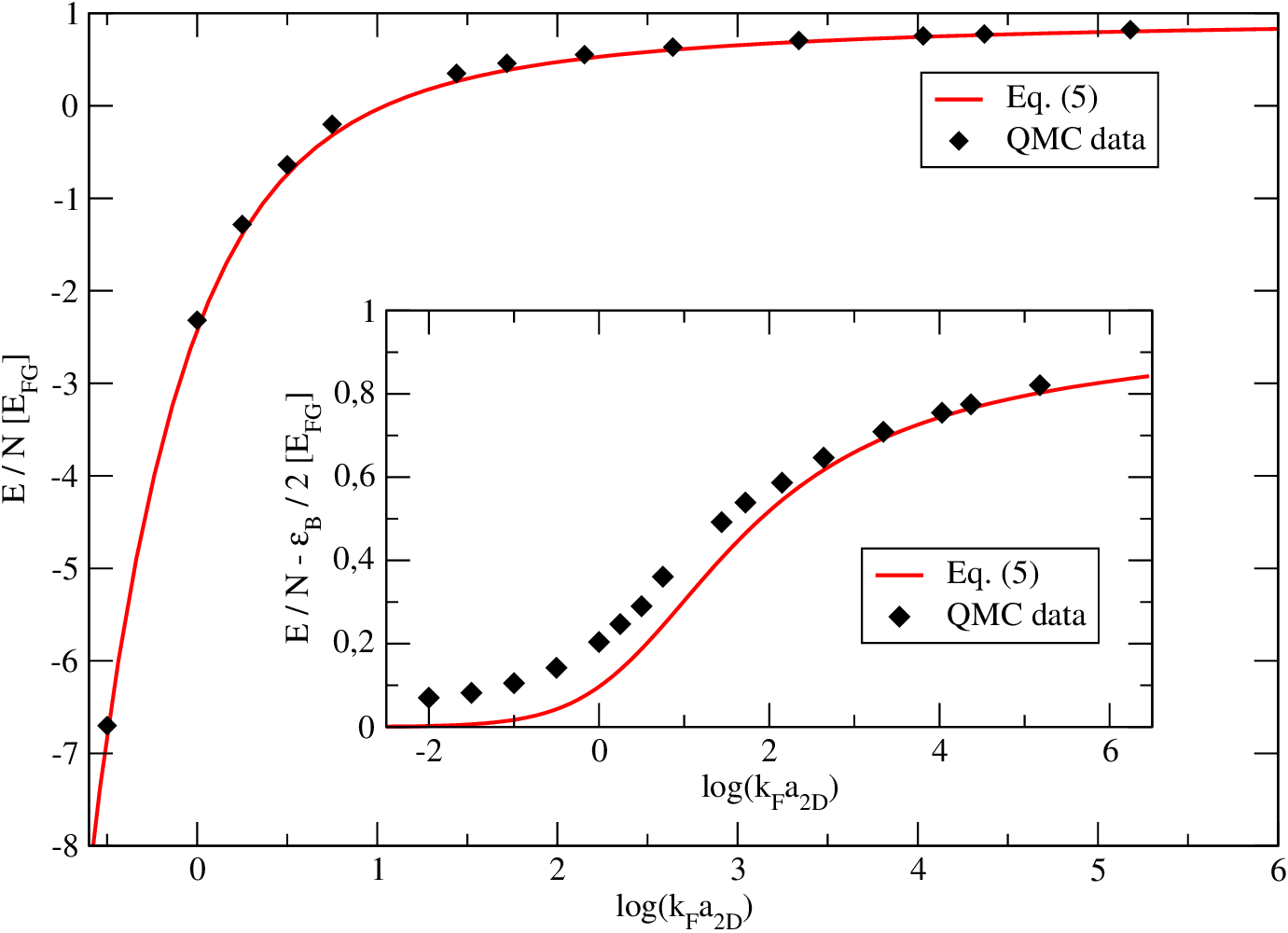}
\caption{Main: Comparison of the energy per particle from Eq. (\ref{E}) as a function of the two-body binding energy $\epsilon_b$ with the Monte-Carlo data from \cite{Bertaina2011}. For the comparison we introduced the 2D scattering length $a_{2D}=2 \hbar e^{-\gamma} / \sqrt{m \epsilon_B}$ \cite{Bertaina2011}.
Inset: Energy per particle with $\epsilon_B / 2$ subtracted.}
\label{fig:E_int}
\end{center}
\end{figure}

Fig. \ref{fig:E_int} shows that our result Eq. (\ref{E}) is in reasonable agreement with Monte-Carlo calculations \cite{Bertaina2011}. Recent experiments \cite{Ong2015, Makhalov2014} confirmed already the Monte-Carlo results. The inset shows the deviations in the molecular limit, following from the fact that interactions between bosonic pairs are neglected. 

Note, that our result Eq. (\ref{E}) is in some sense universal, since it does not depend on the particular shape of the interaction, but only on the two-body binding energy.

From Eq. (\ref{E}) we can now easily derive an expression for the chemical potential. At zero temperature it is defined by $\mu=\partial E/ \partial n$ and we obtain
\begin{align}\label{mu}
\frac{\mu}{\epsilon_F} \approx 1 - \frac{2}{\ln\left[1+\frac{2\epsilon_F}{|\epsilon_b|} \right]}
+ \frac{\frac{2\epsilon_F}{|\epsilon_b|}}{\left(1+\frac{2\epsilon_F}{|\epsilon_b|}\right)
\left(\ln\left[1+\frac{2\epsilon_F}{|\epsilon_b|}\right]\right)^2}.
\end{align}

Knowing the chemical potential and the energy per particle one can also obtain the pressure of the gas at $T=0$. From the Gibbs-Duhem relation $P=n\mu - E/V$  we obtain
\begin{align}\label{P}
\frac{P}{P_{FG}} \approx 1 - \frac{2}{\ln\left[1+\frac{2\epsilon_F}{|\epsilon_b|} \right]}
+ \frac{\frac{\epsilon_F}{|\epsilon_b|}}{\left(1+\frac{2\epsilon_F}{|\epsilon_b|}\right)
\left(\ln\left[1+\frac{2\epsilon_F}{|\epsilon_b|}\right]\right)^2},
\end{align}
where $P_{FG}=\pi n^2 / 2m$ is the pressure of the ideal Fermi gas. 

The pressure has been measured in \cite{Ong2015} and \cite{Makhalov2014}. It was used for a comparison between theory and experiment. Eq. (\ref{E}), Eq. (\ref{mu}) and  Eq. (\ref{P}) can be found already in the appendix of \cite{Ong2015} derived from the polaron model \cite{Klawunn2011} with intuitive arguments. 

\begin{figure}[ht]
\begin{center}
\includegraphics[width=0.44\textwidth,angle=0]{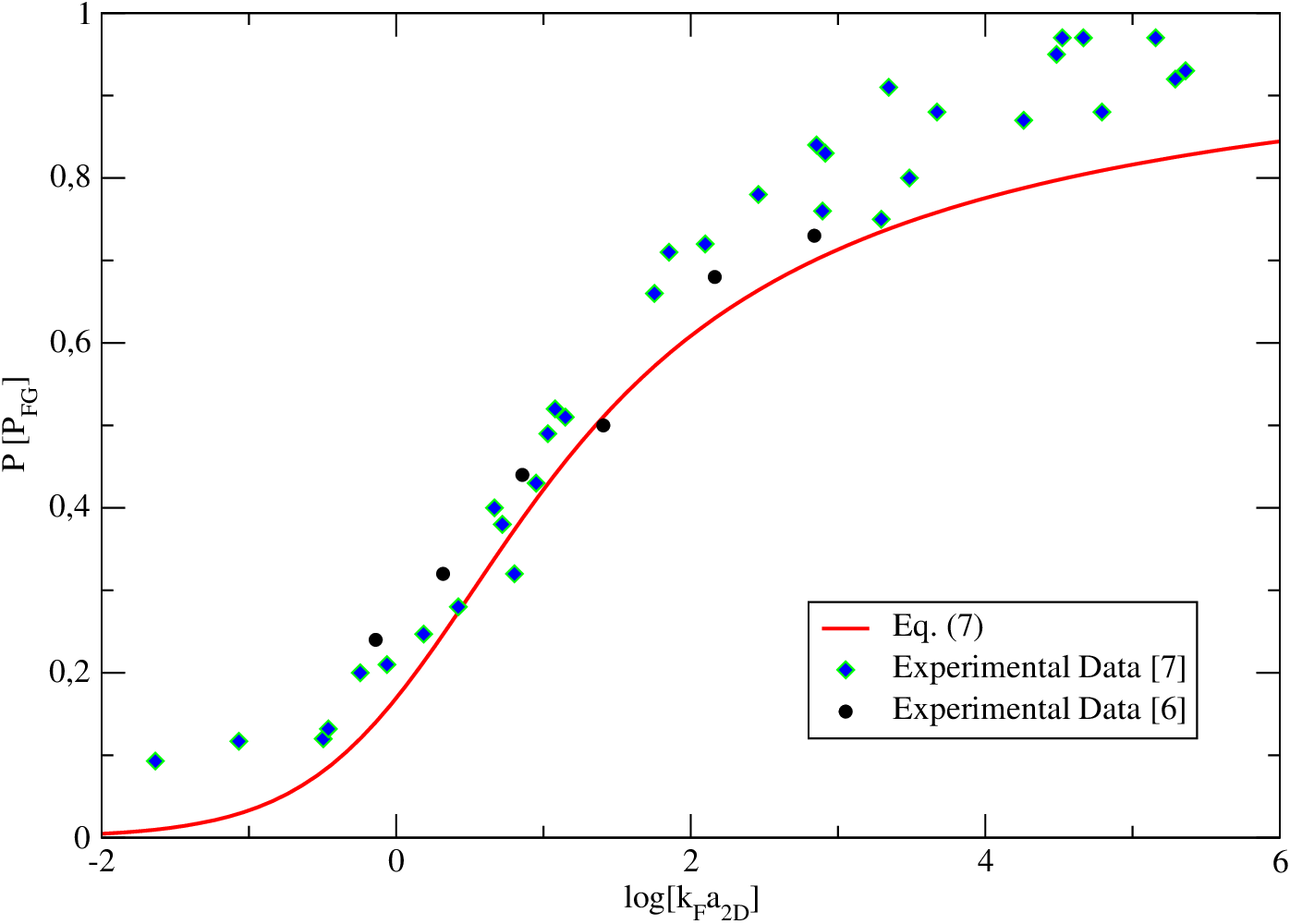}
\caption{Comparison of the pressure of the gas from Eq. (\ref{P}) as a function of $\epsilon_B$ (red line) with the experimental results from \cite{Ong2015} (black dots) and \cite{Makhalov2014}(blue diamonds). For consistency we used the 2D scattering length $a_{2D}=2 \hbar e^{-\gamma} / \sqrt{m \epsilon_B}$ \cite{Bertaina2011} as introduced in the caption of Fig. \ref{fig:E_int}.}
\label{fig:pressure}
\end{center}
\end{figure}

Fig. \ref{fig:pressure} shows that our approach, Eq. (\ref{P}), is in qualitative agreement with the experimental results. Remind, that contrary to the experiments we consider a homogeneous density and zero temperature. Moreover we neglect interactions between bosonic molecules. Therefore an exact quantitative agreement is not expected.


\section{Kosterlitz-Thouless transition temperature}

The two dimensional Bose gas belongs to a class of systems, where topological long-range order occurs at sufficiently low temperatures \cite{KT}. The Kosterlitz-Thouless temperature $T_{KT}$ is defined as the temperature of the phase transition, above which long-range order and hence superfluidity disappear. For the superfluid Bose gas it satisfies the self-consistency equation~\cite{KT}
%
%
\begin{equation}\label{eq-KT} 
k_B T_{KT}=\pi\hbar^2 \rho_s(T_{KT})/2 M^2,
\end{equation}
where $M$ is the boson (dimer) mass ($M=2m$) and $\rho_s$ is the superfluid mass density. 
The superfluid mass density is approximately determined by using 
the expression for the normal density~\cite{Landau9}
\begin{equation}\label{eq-rho}
 \rho_n=\frac{\hbar^2}{2\pi}\int_0^\infty \frac{-\partial n_F}{\partial \epsilon_k} k^3 dk
\end{equation}
and $\rho_s=\rho -\rho_n$,
with $n_F(\epsilon)=(1+e^{\epsilon/T})^{-1}$.

At low temperatures the superfluid density is almost constant, until it jumps at the KT transition temperature to zero. Therefore we approximate $\rho_s(T) \approx  \rho_s(T=0)$ and neglect self-consistency. At zero temperature the derivative of the Fermi distribution function approaches the delta-function $\lim_{T \to 0} (-\partial n_F(\epsilon)/\partial \epsilon)=\delta(\epsilon - \epsilon_F)$. For the crossover $\epsilon_F$ is replaced by the chemical potential $\mu$ from Eq. (\ref{mu}) in the spectrum of fermionic excitations. Then Eq. (\ref{eq-rho}) reduces to $\rho_n / \rho \approx \mu/\epsilon_F$ for $\mu \geq 0$, and zero otherwise. As expected, this expression gives $\rho_n=\rho$ for the non-interacting Fermi gas, where $\mu=\epsilon_F$, such that superfluidity does not occur. With increasing interaction $\mu$, and hence $\rho_n$, tends to zero. Introducing this into the KT-equation (\ref{eq-KT}), we approximately obtain the KT-transition-temperature as a function of the Fermi energy and the two-body binding energy 
\begin{equation}\label{eq-KT-a} 
k_B T_{KT}=\frac{1}{8} \left[ \epsilon_F - \mu(\epsilon_F, \epsilon_B)\right],
\end{equation}
where $\mu \geq 0$ is given by Eq. (\ref{mu}). In the molecular regime Eq. (\ref{eq-KT-a}) yields $k_B T_{KT}=\frac{1}{8} \epsilon_F$ as expected from \cite{Sademelo2006}.

\begin{figure}[ht]
\begin{center}
\includegraphics[width=0.4\textwidth,angle=0]{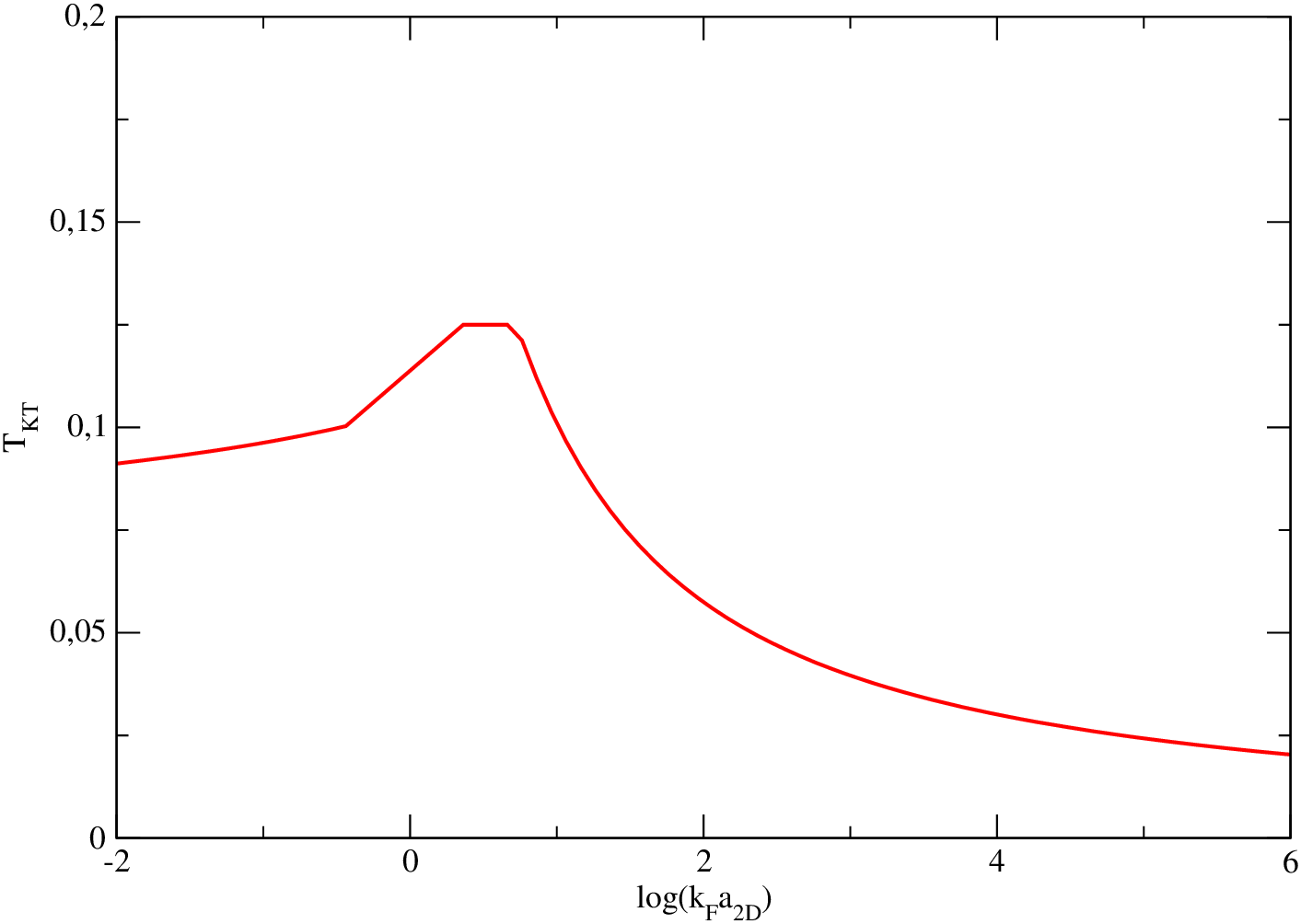}
\caption{KT transition temperature as a function of the two-body binding energy $\epsilon_b$. For $\epsilon_B \leq \epsilon_F$ the figure shows Eq. (\ref{eq-KT-a}) and for the BEC-regime $\epsilon_B > 5 \epsilon_F$ the expression from \cite{Petrov2003}. For consistency we used the 2D scattering length $a_{2D}=2 \hbar e^{-\gamma} / \sqrt{m \epsilon_B}$ \cite{Bertaina2011} as introduced in the caption of Fig. \ref{fig:E_int}.}
\label{fig:KT}
\end{center}
\end{figure}

In our approach collective bosonic excitations are lacking, because interactions between bosonic pairs are neglected. These collective excitations decrease $T_{KT}$ in the molecular regime. In this regime, where $\epsilon_B \gg \epsilon_F$, the transition temperature is basically given by the one derived for the 2D Bose gas \cite{Petrov2003}. Fig. \ref{fig:KT} shows $T_{KT}$ as a function of the binding energy in units of the Fermi energy. In the diagram we used Eq. (\ref{eq-KT-a}) for $\epsilon_B \leq \epsilon_F$ and the expression from \cite{Petrov2003} for $\epsilon_B > 5 \epsilon_F$. Fig. \ref{fig:KT} shows, that somewhere between these two regimes, when $\epsilon_B \approx \epsilon_F$  there should be a maximum. However our approach is not accurate enough to give a quantitative answer, where this maximum appears. 



\section{Conclusion}

In conclusion, we provide an analytical approach for the Fermi gas with attractive interactions. We have derived analytical expressions for the equation of state and the KT transition temperature. 
Our results depend only on the density of the gas via the Fermi energy and the two-particle interaction via the binding energy.
We validated our results by comparison with published results from numerical calculations and experiments.
We believe that this paper can help to increase the understanding of the BCS-BEC crossover in 2D Fermi gases. Further thermodynamic quantities can be derived from our results. Moreover, the analytical expressions could give guidance for future experimental and numerical research on 2D Fermi gases. 
In the next steps one could systematically include the trap, finite temperatures and the interactions between bosonic molecules.
It would be interesting to test the approach and the results from this paper for other types of interactions or ultracold gases in lattices.
The question, whether the approach and the approximations lead to reasonable results in 1D or 3D Fermi gases, could be also studied in future investigations.


\begin{thebibliography}{99}
\bibitem{noteD}
Interestingly this result is independent of dimension. The correction term to the binding energy in $\epsilon_{int}$ is always equal to the energy of the ideal Fermi gas, e.g $\epsilon_F/2$ in 2D and $3/5 \epsilon_F$ in 3D.
\bibitem{note_imbalance}
Note, that, due to our approximations, the same result would be obtained, if one considers the $|\uparrow >$ - atoms as $n/2$ polarons. Therefore our model can also be used to describe imbalanced systems \cite{Ong2015}.
\end{thebibliography}
\end{document}